\documentclass[journal]{IEEEtran}
\ifCLASSINFOpdf
\else
\fi

\usepackage{graphicx}
\usepackage{amssymb,amsmath,epsfig}
\usepackage{latexsym}
\usepackage{graphicx,multirow,threeparttable,float,balance,array,bm,subcaption,caption}
\usepackage{cite}
\usepackage{hyperref}
\usepackage[english]{babel}
\usepackage{diagbox}
\makeatletter
\begin{document}
\title{SpEx: Multi-Scale Time Domain Speaker Extraction Network}
\author{Chenglin~Xu,~\IEEEmembership{Student Member,~IEEE,}
        Wei~Rao,~\IEEEmembership{Member,~IEEE,}
        Eng Siong~Chng,~\IEEEmembership{Senior Member,~IEEE,}
        and~Haizhou~Li,~\IEEEmembership{Fellow,~IEEE}% <-this % stops a space
\thanks{Manuscript received November 27, 2019; revised March 10, 2020 and April 8, 2020; accepted April 8, 2020. Date of publication April 14, 2020 in IEEE/ACM Transactions on Audio, Speech and Language Processing. This work was supported in part by the National Research Foundation Singapore under its AI Singapore Programme (Award No. AISG-100E-2018-006), and in part by Grant No. A1687b0033, Advanced Manufacturing and Engineering, Research, Innovation and Enterprise 2020 Programmatic Fund, Singapore. The work of Haizhou Li is also partly supported by U Bremen Excellence Chairs Program, Germany.  (\textit{Corresponding author: Wei Rao.}) Chenglin Xu and Eng Siong Chng are with the School of Computer Science and Engineering and Temasek Laboratories @ NTU, Nanyang Technological University, Singapore (e-mail:\{xuchenglin, aseschng\}@ntu.edu.sg). Wei Rao and Haizhou Li are with the Department of Electrical and Computer Engineering, National University of Singapore, Singapore (e-mail:\{elerw, haizhou.li\}@nus.edu.sg). Haizhou Li is also with University of Bremen, Germany, and Kriston AI Lab, China.}}

%\thanks{Manuscript received April 1, 2015.}}

%\markboth{Journal of IEEE Audio, Speech and Language Processing Magazine, April~2015}%
%{Shell \MakeLowercase{\textit{et al.}}: Bare Demo of IEEEtran.cls for Journals}



\maketitle

\begin{abstract}
Speaker extraction aims to mimic humans' selective auditory attention by extracting a target speaker's voice from a multi-talker environment. It is common to perform the extraction in frequency-domain, and reconstruct the time-domain signal from the extracted magnitude and estimated phase spectra. However, such an approach is adversely affected by the inherent difficulty of phase estimation. Inspired by Conv-TasNet, we propose a time-domain speaker extraction network (SpEx) that converts the mixture speech into multi-scale embedding coefficients instead of decomposing the speech signal into magnitude and phase spectra. In this way, we avoid phase estimation. The SpEx network consists of four network components, namely \textit{speaker encoder}, \textit{speech encoder}, \textit{speaker extractor}, and \textit{speech decoder}. Specifically, the \textit{speech encoder} converts the mixture speech into multi-scale embedding coefficients, the \textit{speaker encoder} learns to represent the target speaker with a speaker embedding. The \textit{speaker extractor} takes the multi-scale embedding coefficients and target speaker embedding as input and estimates a receptive mask. Finally, the \textit{speech decoder} reconstructs the target speaker's speech from the masked embedding coefficients. We also propose a multi-task learning framework and a multi-scale embedding implementation. Experimental results show that the proposed SpEx achieves 37.3\%, 37.7\% and 15.0\% relative improvements over the best baseline in terms of signal-to-distortion ratio (SDR), scale-invariant SDR (SI-SDR), and perceptual evaluation of speech quality (PESQ) under an open evaluation condition.

\end{abstract}


% \fi
\IEEEpeerreviewmaketitle

\section{Introduction}
\label{sec:introduction}

\IEEEPARstart{T}{}he human brain is able to focus auditory attention on a particular voice by masking out the acoustic background in the presence of multiple talkers and background noises \cite{cherry1953some,getzmann2017switching}. This is called cocktail party effect or cocktail party problem. 

Infants as young as five months have developed the ability to give special attention to their own names \cite{conway2001cocktail}. Behavioral studies have shown that both the abilities to selectively attend to relevant stimuli and to effectively ignore irrelevant stimuli are developed progressively with increasing age across childhood \cite{coch2005event}. These remarkable abilities are implemented with accurate processing of low-level stimulus attributes, segregation of auditory information into coherent voices, and selectively attending to a voice at the exclusion of others to facilitate higher level processing \cite{hill2009auditory}. 

Humans' ability of selective auditory attention has been clearly shown using multi-electrode surface recordings from the auditory cortex \cite{mesgarani2012selective}. Attention is not a static, one way information distillation process. It is believed to be a modulation of focus between the \textit{bottom-up} sensory-driven factors, such as a loud explosion that would attract attention, and the \textit{top-down} task specific goal, such as a flight announcement of one's interest in a busy airport \cite{kaya2017modelling}. The modulation is done rapidly at real-time in response to the input acoustic stimulus and the top-down attention task in the cognitive process. 

Recent physiological studies reveal that such attentional modulation takes place both locally by transforming the receptive field properties of the individual neurons and globally throughout the auditory cortex by rapid neural adaptation, or plasticity, of the cortical circuits \cite{kaya2017modelling}. Computationally, the selective attention to an acoustic stimulus $E(t)$ of interest can be described by a spectro-temporal receptive field, $M(t)$, which acts as a spectro-temporal mask. The modulated response $S(t)$ \cite{kaya2017modelling} to $E(t)$ can be expressed as the  element-wise  multiplication between the stimulus  and the mask, $S(t)= M(t) \otimes E(t)$, where $M(t)$ can be seen as the modulation of the input stimulus by a top-down voluntary focus, or top-down attention.

The top-down attention tasks vary with the application scenarios, for example, the flight announcement from a busy airport, the singing vocal from a music, or the speech of particular speaker from a multi-talker acoustic environment. In this paper, we are interested in how to pay a selective attention to a target speaker, a task which we call speaker extraction. Speaker extraction is highly demanded in real-world applications, such as, hearing aids \cite{wang2017deep}, speech recognition \cite{li2015robust,watanabe2017new,xiao2016study}, speaker verification\cite{rao2019target_is}, speaker diarization\cite{sell2018diarization}, and voice surveillance. A speaker independent speaker extraction system is expected to work for any target speaker unseen during the training, that we call open condition.  

Building on the idea of spectro-temporal receptive field, there have been attempts to perform speaker extraction in frequency-domain through a spectro-temporal mask. 
The studies on computational auditory scene analysis (CASA) \cite{lyon1983computational,meddis1991virtual,ellis1996prediction,seltzer2003harmonic,wang2006computational,hu2007auditory}, non-negative matrix factorization (NMF) \cite{hoyer2004non,cichocki2006new,schmidt2006single,smaragdis2007convolutive,virtanen2007monaural,parry2007incorporating}, and factorial HMM-GMM \cite{virtanen2006speech,kristjansson2006super,stark2011source},  provide invaluable findings for solving the cocktail party problem. With the advent of deep learning, an idea was implemented to optimize the mask of individual speakers with deep recurrent neural networks for source separation of known speakers \cite{huang2015joint}. However, machines have yet to achieve the same  attention ability as humans in the presence of background noise or interference of competing speakers.
The question is how to equip a network the ability to estimate the mask at run-time for a new speaker that is unseen by the system during training.   

The studies on speaker-independent speech separation have seen major progress recently such as deep clustering (DC) \cite{hershey2016deep,isik2016single,wang2018alternative}, deep attractor network (DANet) \cite{chen2017deep,luo2018speaker}, permutation invariant training (PIT) \cite{yu2017permutation,kolbaek2017multitalker,xu2018single,xu2018shifted}, and time-domain audio separation network (TasNet) \cite{luo2018tasnet,luo2018real,luo2018tasnet_arvix,luo2019conv}. Speech separation approaches mimic the human's bottom-up sensory-driven attention. In general, speech separation methods require knowing or estimating the number of speakers in the mixture in advance. However, the number of speakers couldn't always be known in advance in real world applications. Furthermore, speech separation methods may suffer from what is called global permutation ambiguity, where the separated voice for the same speaker may not stick to the same output stream when crossing long pauses or utterances because the separation is done utterance by utterance \cite{luo2019conv}.

\begin{figure}[tb]
\centering
\includegraphics[width=\linewidth]{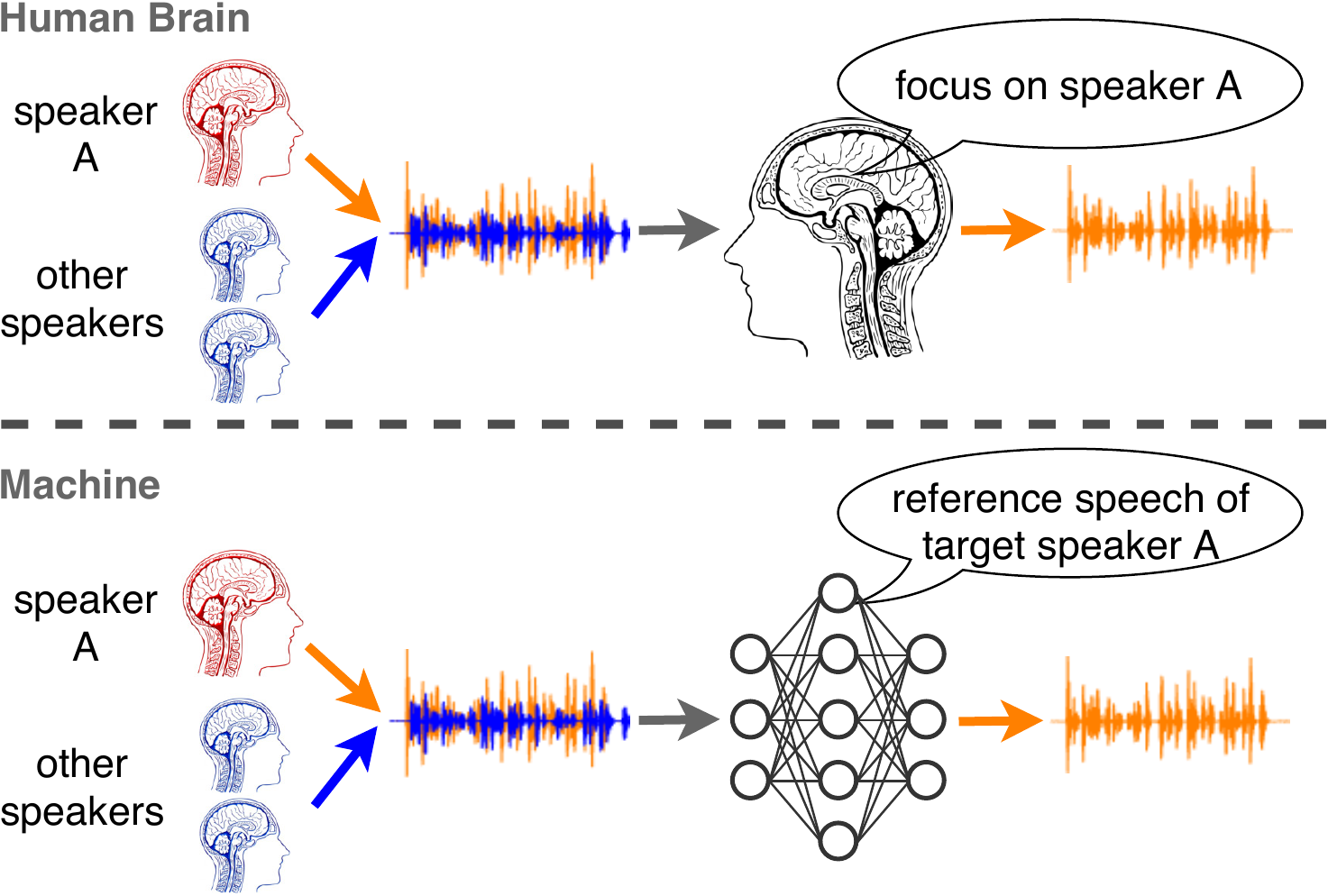}
\caption{Emulating humans' ability of selective auditory attention with speaker extraction network, where a reference speech of target speaker is used to direct the top-down voluntary focus.}
% \vspace{-10pt}
\label{fig:brain}
\end{figure}

Speaker extraction \cite{vzmolikova2017learning,delcroix2018single,wang2018deep,xu2019optimization,xiao2019single,delcroix2019compact,ochiai2019unified,wang2019voicefilter,xu2019time} represents one of the solutions to  the problem of unknown number of speakers and global permutation ambiguity. The idea is to provide a reference speech from a new speaker that is unseen during training. The system then uses such reference speech to direct the attention to the attended speaker, that emulates human's top-down voluntary focus, as shown in Figure \ref{fig:brain}. Such speaker extraction technique is particularly useful when the system is expected to respond to a specific target speaker, for example, in speaker verification \cite{rao2019target_is}, where the reference speech of the target speaker is available through an enrolment process. In the prior work \cite{vzmolikova2017learning,delcroix2018single,wang2018deep,xu2019optimization,xiao2019single,delcroix2019compact,ochiai2019unified,wang2019voicefilter}, a common approach is to perform speaker extraction in frequency-domain, and reconstruct the time-domain signal from the extracted magnitude and estimated phase spectra. Others have also studied complex ratio mask \cite{fu2017complex,williamson2017time,tan2019complex} in speech enhancement. The frequency-domain process relies on short-time Fourier transform (STFT) that faces the windowing effect, and phase estimation problem.

Inspired by Conv-TasNet \cite{luo2018tasnet_arvix,luo2019conv} for speech separation, we propose a novel end-to-end network architecture for speaker extraction (SpEx). SpEx is composed of four network components: a \textit{speech encoder} that encodes the time-domain mixture speech into spectrum-like feature representation that we call embedding coefficients, a \textit{speaker encoder} that learns to represent the target speaker with a speaker embedding, a \textit{speaker extractor} that estimates a receptive mask for the reference speech of the target speaker, and a \textit{speech decoder} that reconstructs the clean speech for the target speaker by modulating the receptive mask with the embedding coefficients of the mixture speech. The SpEx architecture allows the joint training of all these four modules to take place with a multi-task learning algorithm. 

The proposed SpEx is different from our earlier work \cite{xu2019time} where the speaker embedding, i-vector \cite{Dehak&Kenny2011}, is not involved in model training.  It is also different from  \cite{vzmolikova2017learning,delcroix2018single,xu2019optimization,delcroix2019compact} where the speaker embedding is only trained to optimize the signal reconstruction loss.  We will further discuss the difference between SpEx and TasNet in Section \ref{sec:conv_tasnet}. 
This paper makes the following contributions:
\begin{enumerate}
\item We emulate human's ability of selective auditory attention by mimicking the top-down voluntary focus using a speaker encoder.
\item We propose a time-domain solution as an extension to Conv-TasNet from speech separation to speaker extraction, that avoids the phase estimation in frequency-domain approaches.
\item We propose a multi-task learning algorithm to jointly optimize the four network components of SpEx with  an unified training process.
\item We propose a multi-scale encoding and decoding scheme that captures multiple temporal resolutions for improved voice quality.
\end{enumerate}

The rest of the paper is organized as follows. We introduce the novel time-domain speaker extraction network in Section \ref{sec:system}. 
In Section \ref{sec:setup}, we discuss the experimental setup. In Section \ref{sec:results}, we report the experimental results. Section \ref{sec:diss} concludes the study.

% \vspace{-5pt}
\section{Time-domain Speaker Extraction Network}
\label{sec:system}
% \vspace{-5pt}

\begin{figure*}[htb]
\centering
\includegraphics[width=0.8\textwidth]{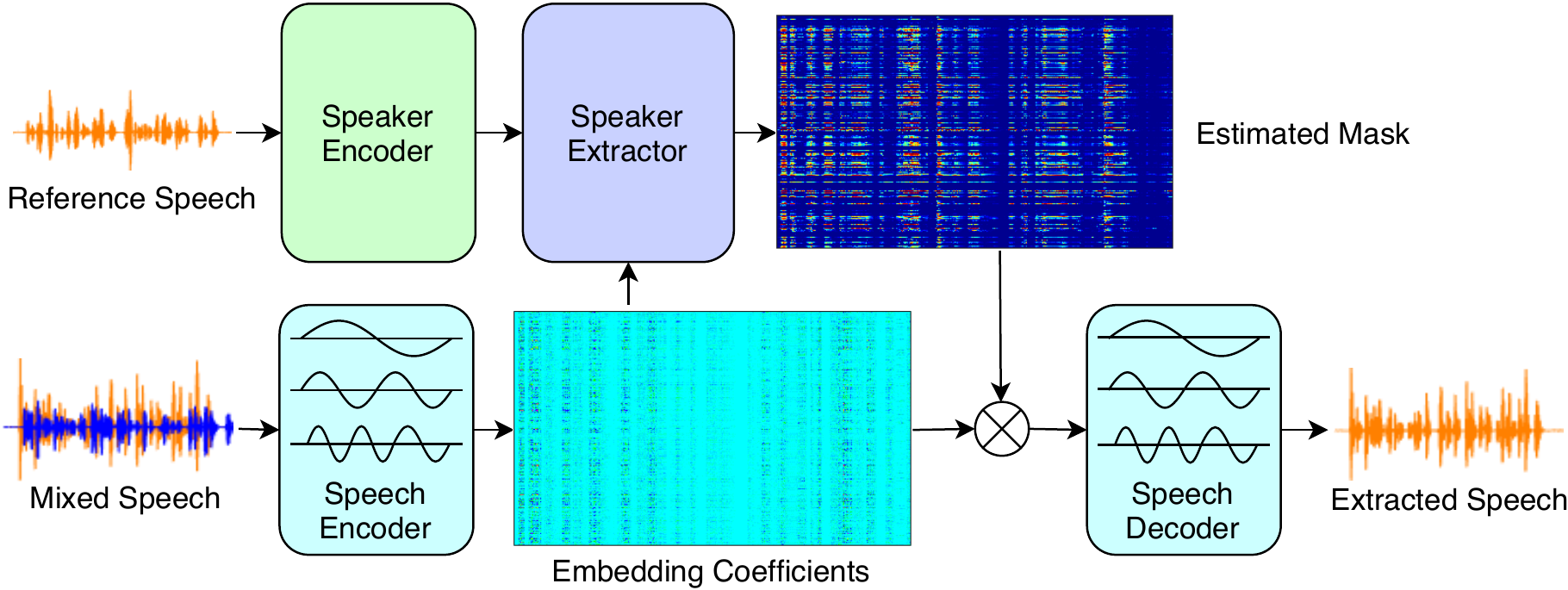}
\caption{The block diagram of a general speaker extraction network, that consists of a speaker encoder (in green), a speech encoder (in cyan), a speaker extractor (in purple), and a speech decoder (in cyan). The network components in  Figure \ref{fig:system} and \ref{fig:SpEx} share the same color codes  for ease of cross reference. The speaker encoder emulates the top-down voluntary focus of cognitive process with the target speaker as the attention task. In the prior work \cite{xu2019time}, the speaker encoder is independently trained. In this paper, it is trained as part of the SpEx network.}
% \vspace{-10pt}
\label{fig:system}
\end{figure*}

A speaker extraction network can be generally described  in Figure \ref{fig:system}. It consists of four network components.  The \textit{speaker encoder} encodes the reference speech $x(t)$ into a speaker embedding, that is the feature representation of the target speaker. The \textit{speech encoder} encodes the time-domain mixture speech $y(t)$ into spectrum or spectrum-like feature representation. The  \textit{speaker extractor} estimates a mask that only lets pass the target speaker's voice. Finally the \textit{speech decoder} reconstructs the time-domain speech signal from the masked spectrum of the mixture speech. From the viewpoint of selective auditory attention, the masked spectrum is called the modulated response \cite{kaya2017modelling}.

\begin{figure*}[htb]
\centering
\includegraphics[width=0.8\textwidth]{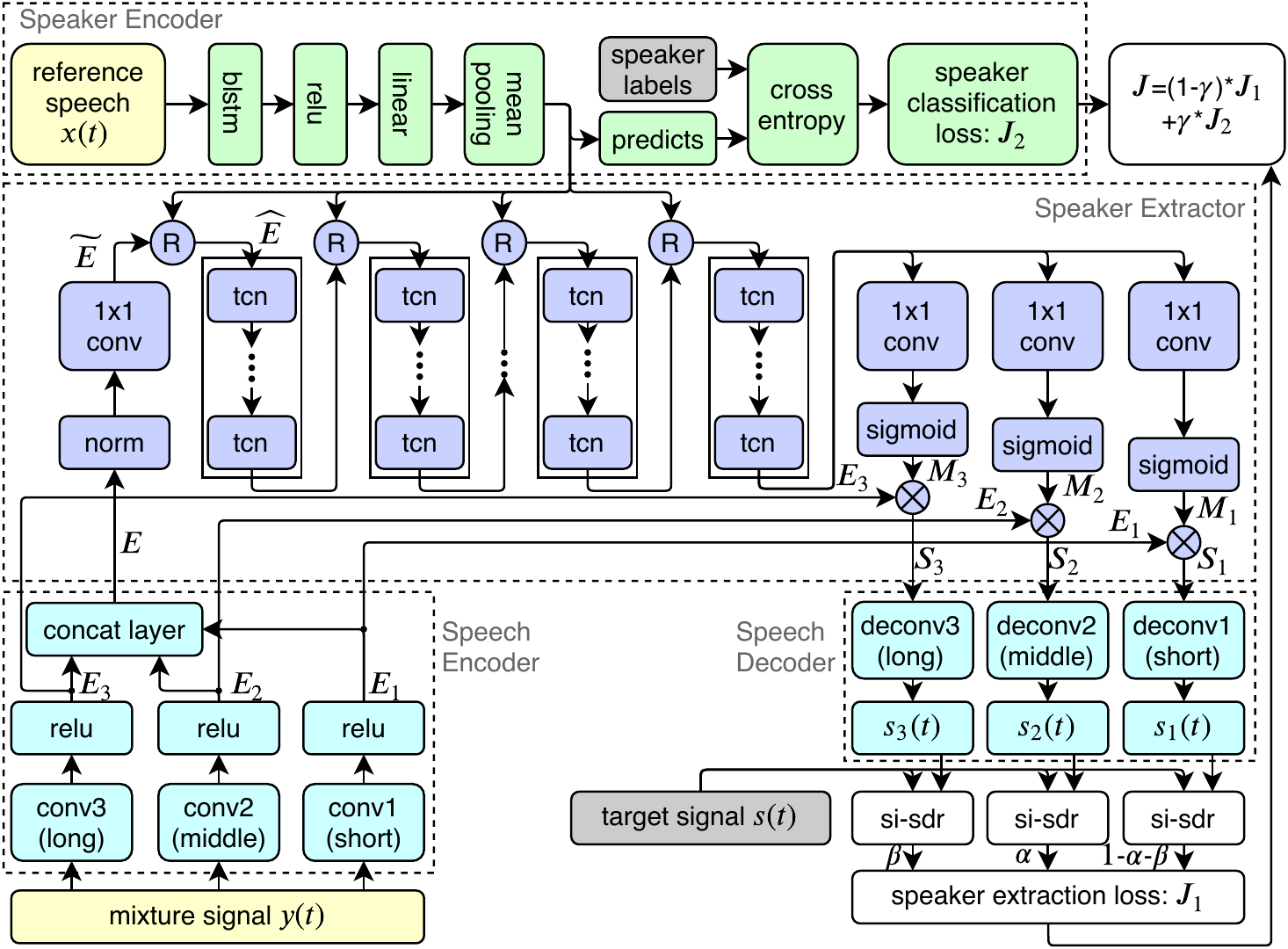}
\caption{The block diagram of the proposed SpEx network, that consists of a speaker encoder (in green), a speech encoder (in cyan), a speaker extractor (in purple), and a speech decoder (in cyan). The network components in  Figure \ref{fig:system} and \ref{fig:SpEx} share the same color codes  for ease of cross reference. \textcircled{\raisebox{-0.9pt}{R}} is an operator that concatenates the speaker vector repeatedly to the intermediate representations of mixture speech along the channel dimension. \textcircled{\raisebox{-0.9pt}{X}} refers to the element-wise multiplication. The ``conv'' and ``deconv'' are 1-D convolutional and de-convolutional operations. ``relu'' and ``sigmoid'' are the rectified linear unit (ReLU) and sigmoid functions. The structure of the ``tcn'' block is similar to Conv-TasNet as shown in Figure \ref{fig:tcn}. The extracted signal $s_1$ is chosen as the ultimate output of the system at run-time inference.}
% \vspace{-10pt}
\label{fig:SpEx}
\end{figure*}

In a frequency-domain implementation, a STFT module serves as the speech encoder that transforms time-domain speech signal into spectrum, with magnitude and phase, while an inverse STFT serves as the speech decoder. Similar to TasNet \cite{luo2018tasnet,luo2019conv}, we opt for a trainable neural network to serve as the speech encoder in time-domain speaker extraction. The speaker encoder is trained to convert time-domain speech signal into spectrum-like embedding, also called embedding coefficients.  The proposed time-domain speaker extraction network (SpEx) is depicted in Figure \ref{fig:SpEx} in detail.

Suppose that a signal $y(t)$ of $T$ samples is the mixture of the target speaker's voice $s(t)$ and $I$ interference voices or background noise $b(t)$. We have,
\begin{equation} \label{eq:discret_signal}
y(t) = s(t) + \sum_{i=1}^I b(t), \quad t=1,...,T
\end{equation}
where $I$ might be any number of interference, and $b(t)$ might be either interference speech or background noise.

During the inference at run-time, given a mixture signal $y(t)$ and a reference speech $x(t)$, the speaker extractor is expected to estimate $\hat{s}(t)$ that is close to $s(t)$ subject to an optimization criterion. 

% \vspace{-5pt}
\subsection{SpEx Architecture}
\label{ssec:SpEx}
% \vspace{-5pt}

\begin{figure}[htb]
\centering
\includegraphics[width=40mm]{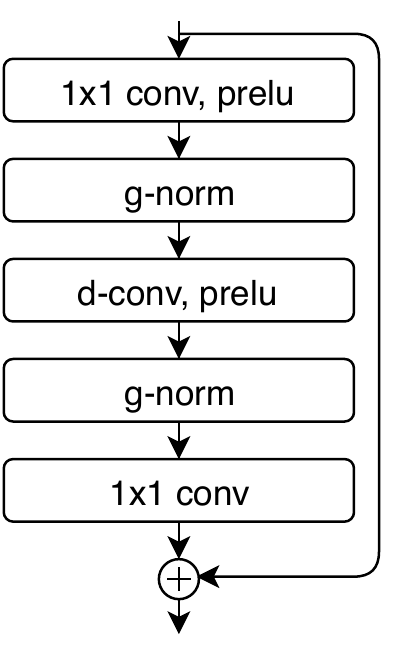}
\caption{The ``tcn'' block is a temporal convolutional network similar to that in Conv-TasNet. \textcircled{\raisebox{-0.9pt}{+}} denotes the residual connection. The ``d-conv'' is depth-wise convolution which forms a depth-wise separable convolution together with the last ``1x1 conv''. ``prelu'' is the parametric rectified linear unit (PReLU). ``g-norm' is the mean and variance  of both time frames and channels scaled by the trainable bias and gain parameters \cite{luo2019conv}. }
% \vspace{-10pt}
\label{fig:tcn}
\end{figure}

\subsubsection{Speaker Encoder}
\label{sssec:embedding_network}

In text-independent speaker recognition, it is common that we represent the speech with a fixed dimensional vector, such as i-vector \cite{Dehak&Kenny2011}, x-vector \cite{snyder2016deep} and other similar feature representations \cite{huang2018angular}, that characterize the voiceprint of a speaker. The model that converts speech samples $x(t)$ into feature representation is called speaker encoder $g(\cdot)$, and the resulting feature representation $g(x)$ is called speaker embedding. 

In \cite{wang2019voicefilter} and \cite{xu2019time}, a speaker encoder is pre-trained independently to extract a d-vector and i-vector for the target speaker. As such speaker encoders are pre-trained independently of speaker extraction systems, they are not optimized directly for speaker extraction purposes. Another idea is to train speaker encoders jointly with the speaker extraction system \cite{vzmolikova2017learning,delcroix2018single,xu2019optimization,delcroix2019compact} with the loss (i.e., mean square error) between the extracted and clean speeches. Such speaker encoders are trained to optimize the signal reconstruction for speaker extraction, however, they do not aim directly to characterize nor discriminate the speakers.

To benefit from the idea of speaker encoder \cite{wang2019voicefilter,xu2019time} and task-oriented optimization \cite{vzmolikova2017learning,delcroix2018single,xu2019optimization,delcroix2019compact}, we propose a multi-task learning algorithm to incorporate the speaker encoder as part of the SpEx network. The speaker encoder is jointly optimized by weighting a cross-entropy loss for speaker classification and a signal reconstruction loss between the extracted and clean speeches for speaker extraction. In practice, we use a bidirectional long-short term memory (BLSTM) to encode the context information of the reference speech into a speaker embedding with a mean pooling layer. In the multi-task learning process, the gradients from both the cross-entropy loss and the signal reconstruction loss are back-propagated to optimize the speaker encoder network. The details of the learning algorithm will be discussed in Section \ref{ssec:multiscale} and \ref{ssec:multitask}.

% \vspace{-5pt}
\subsubsection{Speech Encoder}
\label{sssec:encoder}
% \vspace{-5pt}

There have been studies on how to address the phase estimation problem for frequency-domain methods. One is to optimize the real and imaginary parts separately \cite{fu2017complex,williamson2017time,tan2019complex}, another is to compensate the phase in the training process \cite{gunawan2010iterative,wang2018alternative,wang2018end}. Such attempts have achieved limited successes due to the inexact phase estimation. Similar to Conv-TasNet \cite{luo2018tasnet_arvix,luo2019conv}, we opt for a time-domain approach, that transforms the time-domain mixture signal directly into a feature representation using a convolutional network. 

In a frequency-domain approach, by applying Fourier transform, a speech signal is decomposed into an alternate representation, characterized by sines and cosines. Similarly, in a time-domain approach, we can consider the filters in the convolutional layers as the basis functions in analogy to the sines and cosines in the frequency-domain \cite{reju2007convolution}. The feature representations are considered as the embedding coefficients. After all, the time-domain encoding is different from Fourier transform in that a) the feature representations  don't handle the real-and-imaginary parts separately; b) the basis functions are not pre-defined as sines or cosines, but rather trainable from the data.

The input mixture speech $y(t)\in\mathbb{R}^{1\times T}$ can be encoded to embedding coefficients using a convolutional neural network in a similar way as other end-to-end speech processing systems \cite{sainath2015learning,fu2018end,luo2019conv}. Inspired by \cite{zhu2016learning, multiscale2}, this paper proposes to encode the mixture speech into multi-scale speech embeddings using several parallel 1-D CNNs with $N$ filters each for various temporal resolutions. While the number of multiple scales can vary, we only study three different time scales in this work, without loss of generality. The filters of the parallel 1-D CNNs are of different lengths, $L_1 (short),L_2(middle),L_3(long)$ samples, to cover different window sizes. The CNNs are followed by a rectified linear unit (ReLU) activation function to produce the speech embedding $E=[E_1 E_2 E_3]\in \mathbb{R}^{K\times 3N}$. 

To concatenate the embeddings across different time-scale, we align them by keeping the same stride, $L_1/2$,  across different scales.   With the varying filter lengths, the encoder learns representations in multiple scales, for example, the short window has good resolution at high frequency and long window has high resolution at low frequency. Without trading the temporal resolution with  frequency resolution like in STFT, we encode the time-domain signal into three temporal resolutions in the embedding $E$. The embedding coefficients $E_i$ in each scale consist of a sequence of vectors, $E_{i,k}$, which are defined as,
\begin{equation}
    E_{i,k} = \text{ReLU}(y_{i,k}U_i), \quad k=1,...,K, i=1,2,3
\end{equation}
where $K=2(T-L_1)/L_1+1$, and $y_{i,k}\in \mathbb{R}^{1\times L_i}$ is the $k^{th}$ segment of $y(t)$ that has a window of $L_i$ samples shifting every $L_1/2$ samples. $U_i\in \mathbb{R}^{N\times L_i}$ is also called the encoder basis. 

\subsubsection{Speaker Extractor}
\label{sssec:extractor}

One of the earliest theories of attention is Broadbent's filter model \cite{broadbent1958perception}. In psychoacoustic experiments, the stimuli are first processed according to their physical properties such as color, loudness, and pitch. The selective filters of listeners then allow for certain stimuli to pass through for further processing while other stimuli are rejected. The selective filter can be modelled by a mask that has been well studied in speech separation literature, such as ideal binary mask (IBM) \cite{li2009optimality}, ideal ratio mask (IRM) \cite{narayanan2013ideal}, ideal amplitude mask (IAM) \cite{wang2014training}, wiener-filter like mask (WFM) \cite{erdogan2015phase} and phase sensitive mask (PSM) \cite{erdogan2015phase}. 

In the SpEx framework, the speaker embedding describes the physical properties of the auditory source, a target speaker in this case, as the focus of the attention. The speaker extractor, as shown in Figure \ref{fig:SpEx},  is conditioned on the speaker embedding both during training and at run-time inference to estimate a filter mask, that is referred to as the receptive mask. We obtain the modulated response $S_i$ \cite{kaya2017modelling} for each scale $i=1,2,3$ of the target speaker by applying the receptive mask $M_i$ on the embedding coefficients $E_i$ of the mixture signal in each scale,
\begin{equation} \label{Eq:s1}
\begin{aligned}
    S_i &= M_i \otimes E_i \\
    &= f(E,g(x)) \otimes E_i
\end{aligned}
\end{equation}
where $\otimes$ is an operator for element-wise multiplication. $E$ is the multi-scale embedding coefficients. $f(\cdot)$ and $g(\cdot)$ are the functions representing the speaker extractor and speaker encoder. $x(t)$ is the reference speech of the target speaker to form an attention. 

Specifically, the multi-scale embedding coefficients $E$ are firstly normalized by its mean and variance on channel dimension scaled by the trainable bias and gain parameters. Then, a 1-D CNN with $1\times 1$ kernel size, that is called $1\times 1$ CNN, is applied. The $1\times 1$ CNN with $O$ filters is performed to adjust the number of channels for the inputs and residual path of the subsequent blocks of temporal convolutional network (TCN). In this way, we have the multi-scale embedding coefficients as $\widetilde{E}\in \mathbb{R}^{K\times O}$. At the same time, the speaker embedding vector $g(x)\in \mathbb{R}^{1\times D}$ from the speaker encoder is concatenated repeatedly to $\widetilde{E}$.  The multi-scale embedding coefficients with speaker information are then defined as ${\widehat{E}\in \mathbb{R}^{K\times (O+D)}}$. Similarly, the speaker embedding vector is also concatenated repeatedly with the representations along the subsequent TCN blocks as shown in Figure \ref{fig:SpEx}.

Similar to Conv-TasNet, we stack the TCN blocks by exponentially increasing the dilation factor to capture the long-range dependency of the speech signal. Each TCN block, as shown in Figure \ref{fig:tcn}, applies a dilated depth-wise separable convolution to reduce the number of parameters. The dilated depth-wise separable convolution consists of a dilated depth-wise convolution (``d-conv" in Figure \ref{fig:tcn}) and a following $1\times 1$ CNN with $O$ filters. Since the number of input channels of the TCN block may be different from the number of the filters of the dilated depth-wise convolution, a $1\times 1$ CNN with $P$ filters is applied in advance to turn the number of input channels to $P$. The dilated depth-wise convolution has a kernel size of $1\times Q$, a number of $P$ filters and a dilation factor of $2^{(B-1)}$. $B$ is the number of TCN blocks in a stack. Such a stack is repeated for $R$ times as shown in the speaker extractor in Figure \ref{fig:SpEx}.

To apply the mask $M_i$ on $E_i$,  $M_i$ must have the same dimensions as $E_i$. As the output channels $O$ from the last TCN block may be different from the channels $N$ of the encoded representations $E_i\in\mathbb{R}^{K\times N}$, we  apply one $1\times 1$ CNN to transform the dimension of the output channels from the last TCN block to be same as the encoded representations $E_i\in \mathbb{R}^{K\times N}$. The elements of the mask $M_i\in \mathbb{R}^{K\times N}$ are estimated through a Sigmoid activation function to keep the range within $[0,1]$. Finally, the masked embedding coefficients $S_i\in \mathbb{R}^{K\times N}$ of the target speaker, that are also called the modulated responses \cite{kaya2017modelling}, are estimated by Eq. \ref{Eq:s1}.

\subsubsection{Speech Decoder}
\label{sssec:decoder}

The decoder reconstructs the time-domain speech signal from the modulated responses.  Embedding coefficients at each scale lead to a modulated response. We reconstruct the multi-scale modulated response into time-domain signals ($s_1$, $s_2$, $s_3$) with the decoder bases $V_1\in \mathbb{R}^{N\times L_1}$, $V_2\in \mathbb{R}^{N\times L_2}$, and $V_3\in \mathbb{R}^{N\times L_3}$ through a de-convolutional process. The decoder basis consists of the learned filters during training just as a Fourier basis that is composed of sine and cosine functions.

\subsection{Multi-scale Encoding and Decoding}
\label{ssec:multiscale}

Speech has a rich temporal structure over multiple time scales presenting  phonemic, prosodic and linguistic content \cite{multiscale2}. It was shown that speech analysis of multiple temporal resolutions leads to improved speech recognition performance\cite{multiscale1}.  As shown in Fig. 3, we implement multi-scale encoding in speech encoder and speaker extractor. The speaker encoder first encodes the mixture signal into a multi-scale embedding coefficients $E=[E_1E_2E_3]$. The speaker extractor then estimates multi-scale masks $M_1,M_2,M_3$, and generates the multi-scale modulated responses $S_1,S_2,S_3$. We finally reconstruct the multi-scale modulated responses into time-domain signals $s_1,s_2,s_3$ at multiple scales with the speaker decoder. 

During training, we calculate a multi-scale scale-invariant signal-to-distortion ratio (SI-SDR) loss, defined as $J_1$, that aims to minimize the signal reconstruction error, 
\begin{equation}\label{eq:J1}
    J_1 = -[(1-\alpha-\beta) \rho(s_1, s)+\alpha \rho(s_2,s)+\beta \rho(s_3,s)]
\end{equation}
\noindent{where $\alpha$ and $\beta$ are the weights. $s_1$, $s_2$ and $s_3$ are the signals reconstructed from the modulated responses $S_1$, $S_2$ and $S_3$, respectively. $s$ is the clean speech signal serving as the training target. We use the SI-SDR loss \cite{le2019sdr}, denoted as $\rho(\cdot, \cdot)$, as the measure of reconstruction error.}
\begin{equation}
    {\rho}(\hat{s}, s) = 10\log_{10}(\frac{||\frac{\langle\hat{s}, s\rangle}{\langle s,s\rangle}s||^2}{||\frac{\langle\hat{s}, s\rangle}{\langle s,s\rangle}s-\hat{s}||^2})
\end{equation}
where $\hat{s}$ and $s$ are the extracted and target signals of the target speaker, respectively. $\langle\cdot,\cdot\rangle$ is the inner product. To ensure scale invariance, the signals $\hat{s}$ and $s$ are normalized to zero-mean prior to the SI-SDR calculation. 

The calculation of multi-scale SI-SDR $J_1$ loss is required only during training and not at run-time inference. At run-time inference, we evaluate the quality of the signals reconstructed at multiple scales individually, i.e. $s_1$, $s_2$, $s_3$, and collectively as a weighted summation  $s_w$=(1-$\alpha$-$\beta$)$s_1$+$\alpha s_2$+$\beta s_3$.

\subsection{Multi-task Learning}
\label{ssec:multitask}

We propose to train the speaker encoder together with three other network components as a whole. The speech encoder, speaker extractor, and speech decoder are optimized to minimize the multi-scale SI-SDR loss, while the speaker encoder is optimized with two objective functions, the multi-scale SI-SDR loss and the cross-entropy loss for speaker classification.

The cross-entropy loss $J_2$ for speaker classification is defined as,
\begin{equation} \label{eq:J2}
    J_2 = -\sum_{i=1}^{N_s} p_{i}\log(\hat{p}_i)
\end{equation}
where $N_s$ is the number of speakers in the speaker classification task. $p_i$ is the true class label for speaker $i$ and $\hat{p}_i$ is the predicted speaker probability.

We combine $J_1$ with $J_2$ to optimize the speaker encoder network in a multi-task learning, as $J_1$ and $J_2$ represent two different optimization tasks. With the multi-task learning, the speaker encoder network is trained not only to characterize the unique properties of the target speaker, but also to contribute to the overall speaker extraction objective. The total loss $J$ is a weighted sum of $J_1$ and  $J_2$,
\begin{equation} \label{eq:totoal_loss}
    J = (1-\gamma)J_1+\gamma J_2
\end{equation}

\subsection{Relationship with TasNet}
\label{sec:conv_tasnet}
SpEx network can be seen as an extension to Conv-TasNet \cite{luo2018tasnet_arvix,luo2019conv} from speech separation to speaker extraction applications. A comparison with TasNet framework will help the understanding of SpEx.

BLSTM-TasNet \cite{luo2018tasnet,luo2018real} and Conv-TasNet \cite{luo2018tasnet_arvix,luo2019conv} represent a successful attempt to address the phase estimation problem in frequency-domain speech separation.
The techniques employ an encoder-separator-decoder pipeline, and learn trainable basis functions with a 1-D convolution and de-convolution instead of Fourier series consisting of sine and cosine functions. Speech separation is performed by estimating a mask for each speaker in the mixture using either a BLSTM in BLSTM-TasNet or a fully convolutional neural network (CNN) in Conv-TasNet. Conv-TasNet uses a TCN architecture together with a dilated separable depthwise convolution that represents an effective time-domain implementation. 

The idea of SpEx is similar to that of Conv-TasNet in the sense that the speaker extractor of SpEx is based on the same  TCN architecture\cite{luo2018tasnet_arvix}, and the encoder-extractor-decoder pipeline is inspired by the encoder-separator-decoder pipeline of Conv-TasNet. However, SpEx is also different from Conv-TasNet in the following ways: 

\subsubsection{Top-down  voluntary  focus} SpEx features a speaker encoder as the top-down voluntary focus in selective auditory attention. It learns to single out one voice from the multi-talker mixture by modulating  the input stimulus with a top-down attention. However, Conv-TasNet doesn't employ such a mechanism. It learns to segregate two competing voices by estimating two filtering masks. Just like other speaker extraction techniques, SpEx addresses the issues of global permutation ambiguity and unknown number of speakers that we face in speech separation.

\subsubsection{Multi-task learning}
As Conv-TasNet doesn't involve a speaker encoder, it is trained only to optimize the reconstruction errors, equivalent to the SI-SDR loss in this paper. SpEx adopts a multi-task learning algorithm to jointly optimize all network components, with a cross-entropy loss for speaker classification and a SI-SDR loss for signal reconstruction. The speaker encoder is optimized by the total loss defined in Eq. \ref{eq:totoal_loss}. Such a total loss is different from the prior works, where the speaker extraction systems train the speaker encoder with either speaker classification loss \cite{wang2019voicefilter,xu2019time}, or signal reconstruction loss \cite{vzmolikova2017learning,delcroix2018single,xu2019optimization,delcroix2019compact} as a single task.

\subsubsection{Multi-scale encoding and decoding} The TCN architecture in Conv-TasNet works well for single time scale embedding coefficients \cite{luo2018tasnet_arvix,luo2019conv}. Multi-scale encoding is effective in deep neural networks approach to speech recognition\cite{multiscale2}. We believe that, if the TCN architecture is trained with multi-scale embedding coefficients, it learns to reconstruct the rich temporal structure of speech in greater detail. This will be an interesting study of the TCN architecture.

\section{Experimental Setup} %\vspace{-5pt}
\label{sec:setup}

\subsection{Database}
\label{subsec:data}

We simulated a two-speaker (WSJ0-2mix-extr) and a three-speaker (WSJ0-3mix-extr) mixture databases\footnote{Unlike in speech separation, speaker extraction technique requires a reference speaker to supervise the voluntary attention. We re-organize the well-known WSJ0-2mix and WSJ0-3mix with ``max" data structure by selecting the first chosen speaker as the target speaker, while keeping the mixture speech the same. We rename the simulated database in this work to differentiate from  the original WSJ0-2mix and WSJ0-3mix database. The simulating codes and the best baseline are available at: \url{https://github.com/xuchenglin28/speaker_extraction}. } according to the well-known WSJ0-2mix and WSJ0-3mix \cite{hershey2016deep}. The speech signals are sampled at a sampling rate of 8kHz based on the WSJ0 corpus \cite{garofolo1993csr}. Each database has three datasets: training set ($20,000$ utterances), development set ($5,000$ utterances), and test set ($3,000$ utterances).  

Same as \cite{hershey2016deep}, the training set and development set are generated by randomly selecting two utterances for two-speaker database, and three utterances for three-speaker database, from $50$ male and $51$ female speakers in the WSJ0 ``si\_tr\_s'' set at various signal-to-noise ratio (SNR) between $0$dB and $5$dB. The training set is used for the training of network components, while the development set is for optimizing system configurations.  

Similarly, the utterances from $10$ male and $8$ female speakers in the WSJ0 ``si\_dt\_05'' set and ``si\_et\_05'' set are randomly selected to create the test set. Since the speakers in the test set are excluded from the training and development sets, and the reference speech is not used in any of the speech mixing, the test set is developed for speaker independent evaluation, also called open condition evaluation.  

To include both overlapping and non-overlapping speech in the dataset, we keep the maximum length of the mixing utterances as the length of the mixture. The speaker of the first randomly selected utterance is regarded as the target speaker. At run-time, the speaker extraction process is conditioned on a reference speech from the target speaker. 
 
As the reference speech is randomly selected, the duration of the reference speech varies in training, development and test sets. We call this experimental condition as ``Random". 
In the test set, the average duration of the reference speech is 7.3s with a standard deviation of 2.7s, a maximum length of 19.6s, and a minimum length of 1.6s. The experiments are conducted under this ``Random" condition if not stated otherwise. 

In two-speaker database, we also group the reference speech for the test set into four duration groups, i.e. 7.5s, 15s, 30s and 60s, for the experiment on duration of reference speech, as reported in Section \ref{ssec:duration}. 

While most of the comparative experiments are conducted on the two-speaker database, we also  extend the experiments beyond two-speaker mixture. A three-speaker database is constructed in a similar way as the two-speaker database, except that the duration of the reference speech in the test set is kept as 15s and 60s. In the experiment for three-speaker mixture, we train the SpEx network under three conditions,  two-speaker mixture only,  three-speaker mixture only, and  two-speaker and three-speaker mixture in combination. The trained SpEx systems are then evaluated on the two-speaker and three-speaker mixture test set, respectively. 

The network is optimized by the Adam algorithm \cite{kingma2014adam}. The learning rate begins with $0.001$ and halves when the loss increases on the development set for at least $3$ epochs. Early stopping scheme is applied when the loss increased on the development set for $10$ epochs. The utterances in the training and development set are broken into $4$s segments\footnote{We discard the segments less than 4s or containing only silence for the target speech.}, and the minibatch size is set to $10$.

\subsection{Speaker Encoder}
\label{ssec:speaker}

The speaker encoder in Figure \ref{fig:system} translates the reference speech of the target speech into a top-down voluntary focus that the speaker extractor network can act upon. In Figure \ref{fig:SpEx}, we propose a detailed implementation, that is to repeatedly concatenate the speaker embedding vector with the inputs to TCN blocks. In this paper, we advocate the idea to incorporate the speaker encoder network as an integral part of the SpEx architecture during training and at run-time inference. As a contrastive experiment, we would like to know how such speaker encoder performs differently from a traditional i-vector extractor \cite{Dehak&Kenny2011}. We choose i-vector because it has been one of the most effective techniques for text-independent speaker characterization.

\subsubsection{I-vector Extractor}
\label{sssec:ivec}
An i-vector extractor converts a speech sample into a low-dimension vector. We first train the UBM and total variability matrix with the single speaker (clean) speech from the training and development sets. The acoustic features include 19 MFCCs together with energy, and their 1st- and 2nd-derivatives, followed by cepstral mean normalization \cite{Atal74} with a window size of 3 seconds. The 60-dimensional acoustic features are extracted from a window length of 25ms with a shift of 10ms. A Hamming window is applied. An energy based voice activity detection method is used to remove the silence frames. The i-vector extractor is based on a gender-independent UBM with 512 mixtures and a total variability matrix with 400 total factors.

\subsubsection{Speaker Encoder}
\label{sssec:embedding}
We use the same acoustic features as in the training of i-vector extractor. To leverage the temporal information of the whole reference speech, a BLSTM with $256$ cells in each forward and backward direction is used to capture the speaker information from the acoustic features. A non-linear layer with ReLU activation function with $256$ nodes is followed by the BLSTM. Another linear layer with $400$ nodes followed by a mean pooling is applied to extract the speaker embedding vector, that has the same dimension as the i-vector for fair comparison.

\subsection{Speaker Extraction Pipeline}
\label{subsec:network_config}
The speaker extraction pipeline includes speech encoder, speaker extractor, and speech decoder. The parameters that are quoted in this section have been tuned empirically for the best performance on the development set.

\subsubsection{Speech Encoder}
In the SpEx implementation detailed in Figure \ref{fig:SpEx}, the speech encoder encodes the mixture speech input $Y\in\mathbb{R}^{1\times T}$ into embedding coefficients by three parallel 1-D convolution of $N(=256)$ filters each, followed by a ReLU activation function. To learn multi-scale embeddings with different time resolutions, the three 1-D convolutions had filter lengths of $L_1 (short), L_2 (middle), L_3(long)$ with a stride of $L_1/2$ samples.  $L_1,L_2,L_3$ windows are tuned to cover $20 (2.5ms), 80(10ms), 160(20ms)$ samples in this work.

\subsubsection{Speaker Extractor}
As shown in Figure \ref{fig:SpEx}, a mean and variance normalization with trainable gain and bias parameters is applied to the embedding coefficients $E\in\mathbb{R}^{K\times 3N}$ on the channel dimension, where $K$ is equal to $2(T-L_1)/L_1+1$. A 1x1 convolution linearly transformed the normalized embedding coefficients $E$ to the representations $\widetilde{E}\in\mathbb{R}^{K\times O}$ with $O(=256)$ channels, which determined the number of channels in the input and residual path of the subsequent $1\times 1$ CNN. The number of input channels $P$ and the kernel size $1\times Q$ of each depthwise convolution are set to $512$ and $1\times 3$. $B(=8)$ TCN blocks are formed as a stack and repeated for $R(=4)$ times. %(Haizhou: in the table, B is not always 8, R is not always 4, O is not always 256)

\subsubsection{Speech Decoder}
The speech decoder in Figure \ref{fig:SpEx} reconstructs the time-domain speech signal ($s_1$, $s_2$, $s_3$) from the modulated responses ($S_1$, $S_2$, $S_3$) through a de-convolution process. The filter in the de-convolution has the same configuration as that in the speech encoder, where the number of filters ($N$) is equal to $256$ and the filter lengths ($L_1,L_2,L_3$) are tuned to be $20 (2.5ms), 80(10ms), 160(20ms)$ samples.

\subsection{Reference Baselines}
\label{subsec:baselines}

We select $4$ systems that represent the recent advances in single channel target speaker extraction as the baselines, and implement all of them for benchmarking. The baseline systems belong to the Speaker Beam Frontend (SBF) \cite{delcroix2018single} family, which demonstrates state-of-the-art performance of frequency-domain speaker extraction techniques on databases that are similar to this paper. 

\begin{itemize}
    \item SBF-IBM \cite{delcroix2018single}: This architecture adopts a speaker adaptation layer in a context adaptive deep neural network (CADNN) \cite{delcroix2016context} to track the target speaker from the input mixture in the speaker extraction. The weights in the adaptation layer are learned from a target speaker's enrolled speech in the speaker embedding network. IBM is used to calculate the mask approximation loss as the objective function.
    \item SBF-MSAL \cite{delcroix2018single}: This architecture replaces the IBM objective function in SBF-IBM with a magnitude spectrum approximation loss (MSAL) to directly minimize the signal reconstruction error. It is reported that SBF-MSAL outperforms SBF-IBM.
    \item SBF-MTSAL \cite{xu2019optimization}: This architecture replaces the IBM objective function in SBF-IBM with a magnitude and temporal spectrum approximation loss (MTSAL), in which a temporal constraint is incorporated to ensure the temporal continuity of the output signal. It is reported that SBF-MTSAL outperforms SBF-MSAL.
    \item SBF-MTSAL-Concat \cite{xu2019optimization}: This architecture adopts a BLSTM as the speaker encoder to capture long range speaker characteristics. While the speaker encoder of SBF-MTSAL-Concat is similar to that of SpEx, it is trained only using the magnitude and temporal spectrum approximation loss without multi-task learning. No speaker classification loss is investigated. Nonetheless, it is reported that SBF-MTSAL-Concat outperforms all the above three SBF variations. 
\end{itemize}

\subsection{Evaluation Metrics}
\label{subsec:eval_metrics}
We follow the same evaluation metrics in the speaker extraction literature \cite{xu2019optimization} for ease of comparison. They are the signal-to-distortion ratio (SDR) \cite{vincent2006performance} and perceptual evaluation of speech quality (PESQ) \cite{rix2001perceptual}. We also include SI-SDR \cite{le2019sdr}, because SI-SDR is more suitable and robust for single channel speech separation or extraction than SDR. Since the speaker extraction aims to improve the speech quality and intelligibility, the subjective evaluation of A/B preference test is also conducted to evaluate the perceptual quality of the extracted speech by humans.

\section{Results}
\label{sec:results}
We report the results of 10 experiments in two groups. The first 9 experiments are carried out on the two-speaker mixture database, while the last experiment is on the three-speaker mixture database. 

\subsection{Experiments on Two-speaker Mixture}

\subsubsection{Frequency-domain vs. Time-domain}
\label{subsec:freq_time}

In this experiment, we would like to compare between two processing paradigms, the frequency-domain and the proposed time-domain methods. For frequency-domain implementation, we adopt STFT and inverse STFT as the speech encoder and decoder in Figure \ref{fig:system}    respectively. For time-domain implementation, we adopt the speech encoder and decoder proposed in Section II.  In both systems, we adopt i-vector extractor as the speaker encoder. As the i-vector extractor is trained independently from the speaker extraction pipeline, this comparison is focused on frequency-domain and time-domain speaker extraction pipeline. As the frequency-domain method uses a fixed short-time window of 256 samples, the time-domain systems are also implemented with a single short-time window, or single scale as opposed to multi-scale as discussed in Section \ref{ssec:SpEx}, for fair comparison.

We observe from Table \ref{tbl:freq_time}, that the time-domain speaker extraction systems (System 2-13) consistently outperform the frequency-domain counterpart (System 1), especially when time-domain systems have fewer than or roughly the same number of parameters as the  frequency-domain system.

The results clearly show the advantage of the trainable speech encoder and decoder over the static STFT and inverse STFT in the frequency-domain. We consider that the better performance is attributed to the use of embedding coefficients in place of magnitude and phase spectra in the process, that avoids the need of phase estimation. 

\begin{table*}[t]
\centering \caption{ SDR (dB), SI-SDR(dB) and PESQ in a comparative study between frequency-domain and time-domain under open condition. $L_1$ is the filter length of the convolution in the speech encoder for single scale in this experiment. N, O, P, Q, B, R are the parameters of the extractor defined in Section \ref{sssec:extractor}. In the frequency-domain implementation, we use the phase spectrum from the original mixture speech to reconstruct the speech signal. ``\#Paras" indicates the total number of parameters in the network. i-vector is used as feature representation of reference speaker.} 
\centerline{
\footnotesize
\begin{tabular}{|c|*{12}{c|}}
\hline
System & Domain & N & $L_1$ & O & P & Q & B & R & \#Paras & SDR & SI-SDR  & PESQ  \\
\hline
\hline
1 & Frequency & - & 256 & 256 & 512 & 3 & 8 & 4 & 9.0M & 10.3 & 9.9 & 2.85 \\ \hline
2 & \multirow{12}{*}{Time} & 128 & 20 & 128 & 128 & 3 & 8 & 4 & 1.3M & 12.3 & 11.7 & 2.85 \\
3 & & 128 & 20 & 128 & 128 & 3 & 8 & 5 & 1.7M & 12.0 & 11.2 & 2.82 \\
4 & & 512 & 20 & 128 & 256 & 3 & 8 & 4 & 2.6M & 12.4 & 11.6 & 2.83 \\
5 & & 512 & 20 & 128 & 512 & 3 & 8 & 3 & 3.7M & 11.7 & 10.9 & 2.78 \\
6 & & 256 & 20 & 256 & 256 & 3 & 8 & 4 & 4.7M & 12.6 & 11.9 & 2.88 \\
7 & & 512 & 20 & 128 & 512 & 3 & 8 & 4 & 4.9M & 12.9 & 12.1 & 2.89 \\
8 & & 256 & 20 & 256 & 256 & 3 & 9 & 4 & 5.2M & 12.8 & 12.2 & 2.89 \\
9 &  & 256 & 20 & 256 & 512 & 3 & 8 & 4 & 9.0M & \textbf{13.1} & \textbf{12.4} & 2.92 \\
10 & & 256 & 40 & 256 & 512 & 3 & 8 & 4 & 9.1M & 12.7 & 11.9 & 2.90 \\
11 & & 256 & 80 & 256 & 512 & 3 & 8 & 4 & 9.1M & 13.0 & 12.4 & 2.93 \\
12 & & 256 & 160 & 256 & 512 & 3 & 8 & 4 & 9.1M & 12.2 & 11.5 & 2.88 \\
13 & & 256 & 256 & 256 & 512 & 3 & 8 & 4 & 9.2M & 12.8 & 12.2 & \textbf{2.94} \\
\hline
\end{tabular}} %\vspace{-10pt}
\label{tbl:freq_time}
\end{table*}

\subsubsection{Single-scale vs. Multi-scale}

\begin{table*}[t]
\centering \caption{ SDR (dB), SI-SDR (dB) and PESQ in a comparative study between single-scale and multi-scale under open condition. $L_1$, $L_2$ and $L_3$ are the various filter lengths of convolutions in the speech encoder. N (256), O (256), P (512), Q (3), B (8), R (4) are the parameters of the extractor defined in Section \ref{sssec:extractor}. $\alpha$ and $\beta$ are the weights defined in the multi-scale SI-SDR loss $J_1$ in Eq. \ref{eq:J1}. ``\#Paras" indicates the total number of parameters in the network. $s_w$=(1-$\alpha$-$\beta$)$s_1$+$\alpha s_2$+$\beta s_3$ denotes the weighted summation of the reconstructed signal. The number of parameters during evaluation is less than that of training when only picking $s_1$ as the reconstructed signal. i-vector is used as feature representation of reference speaker.} 
\centerline{
\footnotesize
\begin{tabular}{|c|*{13}{c|}}
\hline
\multirow{2}{*}{System} & \multirow{2}{*}{$L_1$} & \multirow{2}{*}{$L_2$} & \multirow{2}{*}{$L_3$} & \multirow{2}{*}{$\alpha$} & \multirow{2}{*}{$\beta$} & \multicolumn{2}{c|}{Single vs Multiple Scale} & Loss & Reconstructed & \multirow{2}{*}{\#Paras} & \multirow{2}{*}{SDR} & \multirow{2}{*}{SI-SDR}  & \multirow{2}{*}{PESQ}  \\ \cline{7-8}
& & & & & & Speech Encoder & Speech Decoder & Function & Signal & & & & \\
\hline
\hline
9 & 20 & - & - & - & - & single & single & $\rho(s_1,s)$ & $s_1$ & 9.0M & 13.1 & 12.4 & 2.92 \\ \hline
14 & 20 & 80 & 160 & - & - & multiple & single & $\rho(s_1,s)$ & $s_1$ & 9.2M & 13.6 & 13.0 & 3.00 \\
15 & 20 & 80 & 160 & 0.05 & 0.05 & multiple & multiple & $J_1$ & $s_1$ & 9.4M & 12.6 & 11.9 & 2.84 \\
16 & 20 & 80 & 160 & 0.10 & 0.10 & multiple & multiple & $J_1$ & $s_1$ & 9.4M & \textbf{13.9} & \textbf{13.3} & 3.00 \\
17 & 20 & 80 & 160 & 0.20 & 0.20 & multiple & multiple & $J_1$ & $s_1$ & 9.4M & 13.2 & 12.6 & 2.94 \\
18 & 20 & 80 & 160 & 0.33 & 0.33 & multiple & multiple & $J_1$ & $s_1$ & 9.4M & 12.5 & 11.8 & 2.86 \\
19 & 20 & 80 & 160 & 0.10 & 0.05 & multiple & multiple & $J_1$ & $s_1$ & 9.4M & 12.4 & 11.4 & 2.84 \\
20 & 20 & 80 & 160 & 0.20 & 0.10 & multiple & multiple & $J_1$ & $s_1$ & 9.4M & 13.1 & 12.4 & 2.93 \\
21 & 20 & 80 & 160 & 0.30 & 0.20 & multiple & multiple & $J_1$ & $s_1$ & 9.4M & 13.0 & 12.4 & 2.89 \\ \hline
22 & 20 & 80 & 160 & 0.10 & 0.10 & multiple & multiple & $J_1$ & $s_2$ & 9.4M & 12.2 & 11.4 & \textbf{3.01} \\
23 & 20 & 80 & 160 & 0.10 & 0.10 & multiple & multiple & $J_1$ & $s_3$ & 9.4M & 12.1 & 11.4 & 3.00 \\
24 & 20 & 80 & 160 & 0.10 & 0.10 & multiple & multiple & $J_1$ & $s_w$ & 9.4M & 13.9 & 13.3 & 3.00 \\
\hline
\end{tabular}} %\vspace{-10pt}
\label{tbl:mscmo}
\end{table*}

In this experiment, we would like to validate the idea of multi-scale speech embedding. We continue to use i-vector extractor as the speaker encoder. From the experiments reported in Table \ref{tbl:freq_time}, we observe that systems of more parameters perform better. By varying the filter length of the convolution layer in the speech encoder from System 9-13, we observe that the change of time-frequency resolution of the embedding coefficients has an impact on the system performance. The best SDR is achieved as 13.1dB with a filter length of 20 samples ($2.5ms$). The best SI-SDR is 12.4dB with the filter length of 20 samples ($2.5ms$) and 80 samples ($10ms$). The best PESQ is 2.94 with a filter length of 256 samples ($32ms$). This finding is similar to that in speech recognition experiment \cite{multiscale2} .

To benefit from the different time-frequency resolutions, we propose to have three 1-D CNNs with different filter length, short, middle, and long, in the speech encoder. The speaker extractor and speech decoder are also extended to be compatible for the  multi-scale speech embedding, as shown in Figure \ref{fig:SpEx}. The speaker extractor estimates the mask for the target speaker at each scale. The speech decoder reconstructs the time-domain signal for each scale with the modulated response. 

We explore different system configurations that are summarized in System 14-24 of Table \ref{tbl:mscmo}. Comparison between System 9 and System 14 shows that the multi-scale speech encoder achieves better performance than single-scale speech encoder. If the speech decoder has multiple outputs with the multi-scale speech embeddings, we could optimize the SpEx network with a weighted multi-scale SI-SDR loss, as defined in Eq. \ref{eq:J1}. With multi-scale speech encoder and decoder, the best performances of the SDR, SI-SDR and PESQ are achieved at $13.9$dB, $13.3$dB and $3.00$ when the weights $\alpha$ and $\beta$ in Eq. \ref{eq:J1} are tuned to be $0.10$ and $0.10$. Comparing with the single-scale system, the performance of the multi-scale SpEx improves the SDR of $6.1\%$, the SI-SDR of $7.3\%$, and the PESQ of $2.7\%$. Comparisons between System 16 and System 22-24 present that the best performance is achieved by picking the output stream $s_1$ with short window (high temporal resolution). By only picking the reconstructed signal $s_1$ instead of a weighted summation ($s_w$=(1-$\alpha$-$\beta$)$s_1$+$\alpha s_2$+$\beta s_3$), the number of parameters during evaluation is less than that during training. 

\subsubsection{I-vector vs. Speaker Embedding}
\label{ssec:ivector}

We have observed that the i-vector is effective in speaker characterization for both single-scale and multi-scale speaker extraction networks as reported in Tables \ref{tbl:freq_time} and \ref{tbl:mscmo}. We note that the i-vector is extracted independently of the speaker extraction network. In this experiment, we would like to replace the i-vector extractor with the speaker encoder. The speaker encoder is trained jointly with other components of the network using both the cross-entropy loss for speaker classification and the multi-scale SI-SDR loss for speaker extraction as System 25 to 31 in Table \ref{tbl:joint_embed}. 

We obtain the best SDR and SI-SDR of $15.1$dB and $14.6$dB when the weight for the sub-loss of the cross-entropy is tuned to be $0.2$. Comparing with the i-vector based system (System 16 in Table \ref{tbl:mscmo}), we observe that the joint optimization of the speaker encoder and the speaker extraction pipeline (System 27 in Table \ref{tbl:joint_embed}) with multi-task learning achieves relative improvements of  $8.6\%$ in terms of SDR,  $9.8\%$ in terms of SI-SDR,  $4.7\%$ in terms of PESQ. As the SpEx network with joint optimization (Figure 3) achieves the best performance, we use the configuration hereafter. 

\begin{table*}[t]
\centering \caption{ SDR (dB), SI-SDR(dB) and PESQ in a comparative study between i-vector and speaker embedding as feature representations of reference speaker under open condition. $L_1$ (20), $L_2$ (80) and $L_3$ (160) are the various filter lengths of convolutions in the speech encoder. N (256), O (256), P (512), Q (3), B (8), R (4) are the parameters of the extractor defined in Section \ref{sssec:extractor}. $\alpha$ and $\beta$ are the weights defined in the multi-scale SI-SDR loss $J_1$ in Eq. \ref{eq:J1}. $\gamma$ is the weight of multi-task learning defined in Eq. \ref{eq:totoal_loss}. ``MTL'' indicates whether the multi-task learning is applied. ``\#Paras" indicates the total number of parameters in the network. $s_w$=(1-$\alpha$-$\beta$)$s_1$+$\alpha s_2$+$\beta s_3$ denotes the weighted summation of the reconstructed signal. The number of parameters during evaluation is less than training when only picking $s_1$ as the output. } 
\centerline{
\footnotesize
\begin{tabular}{|c|*{12}{c|}}
\hline
\multirow{2}{*}{System} & \multirow{2}{*}{$\alpha$} & \multirow{2}{*}{$\beta$} & \multirow{2}{*}{$\gamma$} & \multirow{2}{*}{Speaker Characterization} & Speaker Encoder & \multirow{2}{*}{MTL} & Loss & Reconstructed & \multirow{2}{*}{\#Paras} & \multirow{2}{*}{SDR} & \multirow{2}{*}{SI-SDR}  & \multirow{2}{*}{PESQ}  \\ 
& & & & & Joint Optimization & & Function & Signal & & & & \\
\hline
\hline
16 & 0.1 & 0.1 & - & i-vector & no & no & $J_1$ & $s_1$ & 9.4M & 13.9 & 13.3 & 3.00 \\ \hline
25 & 0.1 & 0.1 & - & speaker embedding & yes & no & $J_1$ & $s_1$ & 10.8M & 14.2 & 13.7 & 3.04 \\
26 & 0.1 & 0.1 & 0.1 & speaker embedding & yes & yes & $J$ & $s_1$ & 10.8M & 15.0 & 14.6 & \textbf{3.15} \\
27 & 0.1 & 0.1 & 0.2 & speaker embedding & yes & yes & $J$ & $s_1$ & 10.8M & \textbf{15.1} & \textbf{14.6} & 3.14 \\
28 & 0.1 & 0.1 & 0.3 & speaker embedding & yes & yes & $J$ & $s_1$ & 10.8M & 14.3 & 13.8 & 3.03 \\ \hline
29 & 0.1 & 0.1 & 0.2 & speaker embedding & yes & yes & $J$ & $s_2$ & 10.8M & 12.8 & 12.2 & 3.15 \\
30 & 0.1 & 0.1 & 0.2 & speaker embedding & yes & yes & $J$ & $s_3$ & 10.8M & 12.8 & 12.2 & 3.15 \\
31 & 0.1 & 0.1 & 0.2 & speaker embedding & yes & yes & $J$ & $s_w$ & 10.8M & 14.9 & 14.4 & 3.13 \\
\hline
\end{tabular}} %\vspace{-10pt}
\label{tbl:joint_embed}
\end{table*}

\subsubsection{Benchmark against the Baselines}

We compare the SpEx network as illustrated in Figure \ref{fig:SpEx} with four competitive baselines \cite{delcroix2018single,xu2019optimization}. As can be seen in  Table \ref{tbl:methods_comp}, the SpEx network shows $37.3\%$, $37.7\%$ and $15.0\%$ relative improvements over the best baseline, SBF-MTSAL-Concat, in terms of SDR, SI-SDR and PESQ under the open condition. 

The time-domain speaker extraction architecture has shown three clear advantages over its frequency-domain counterparts. 

(1) Because the SpEx network doesn't decompose the speech signal into magnitude and phase spectra, it avoids inexact phase estimation.

(2) The SpEx network benefits from the long-range dependency of the speech signal captured by the stacked dilated depth-wise separable convolution with a manageable number of parameters. Without the recurrent connection, the SpEx method can be easily parallelized for fast training and inference. 

(3) The SpEx network takes advantage of  multi-scale speech embedding to have a good coverage of time-frequency resolution in the encoding, which doesn't have to trade time resolution with frequency resolution like in short-time frequency analysis.

\begin{table}[t]
\centering \caption{ SDR (dB), SI-SDR(dB) and PESQ of extracted speech for the proposed SpEx network and other 4 competitive baseline systems  under open condition. ``Mixture'' refers to original input mixture with zero effort. ``\#Paras'' means the number of parameters of the model.} 
\centerline{
\footnotesize
\begin{tabular}{|r|*{5}{c|}}
\hline
Methods & \#Paras & SDR & SI-SDR  & PESQ  \\
\hline
\hline
Mixture & - & 2.6 & 2.5 & 2.31 \\
\hline
SBF-IBM \cite{delcroix2018single} & 19.3M & 6.5 & 6.3 & 2.32 \\ 
SBF-MSAL \cite{delcroix2018single} & 19.3M & 9.6 & 9.2 & 2.64 \\
SBF-MTSAL \cite{xu2019optimization} & 19.3M & 9.9 & 9.5 & 2.66 \\ 
SBF-MTSAL-Concat \cite{xu2019optimization} & 8.9M & 11.0 & 10.6 & 2.73 \\ \hline
SpEx & 10.8M & \textbf{15.1} & \textbf{14.6} & \textbf{3.14} \\
\hline
\end{tabular}} %\vspace{-10pt}
\label{tbl:methods_comp}
\end{table}

\begin{figure}[t]
\begin{center}
\includegraphics[width=\linewidth]{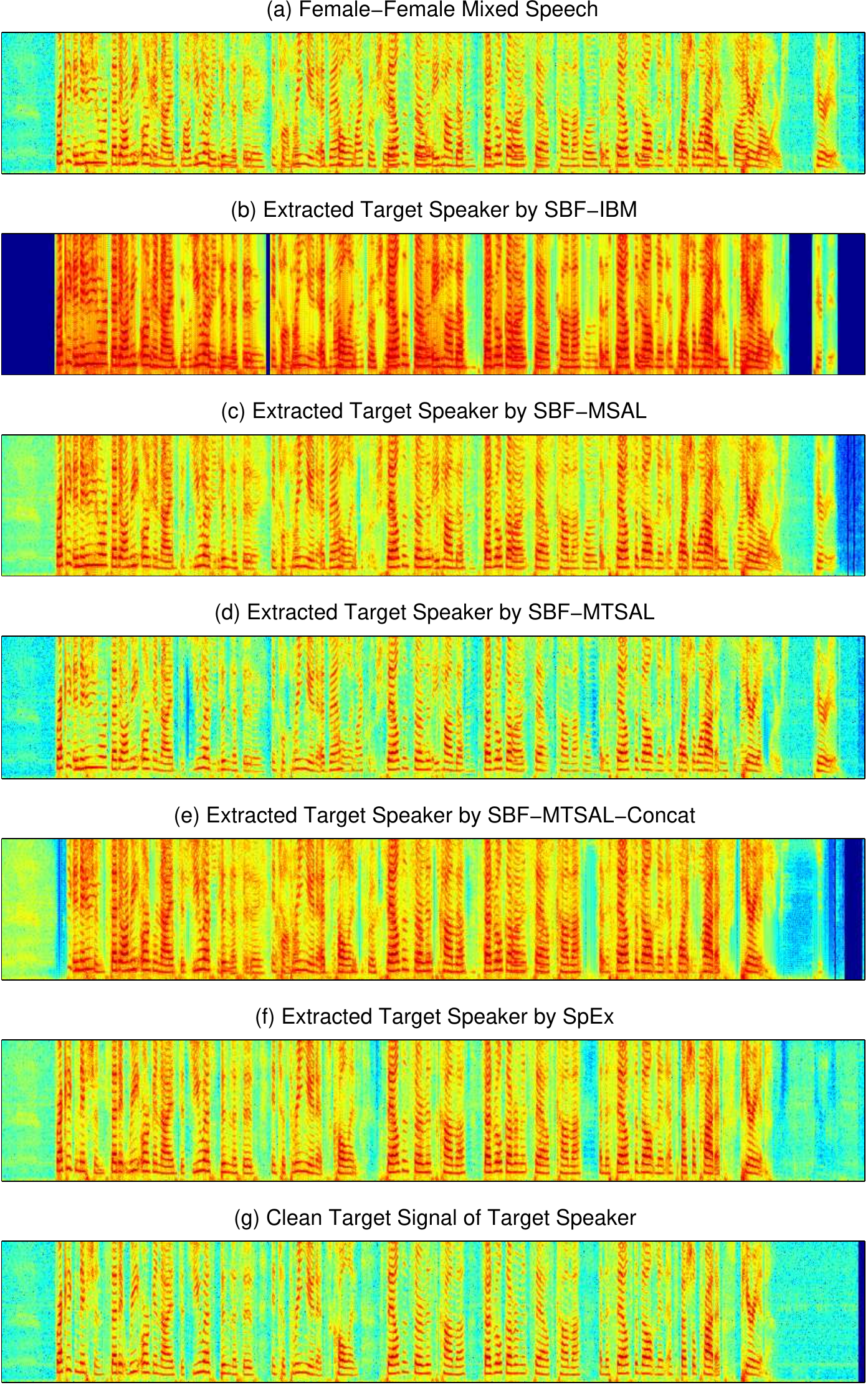} %\vspace{-5pt}
\caption{The log magnitude spectra of a female-female mixture, its extracted speech for target speaker by the four baselines, the proposed SpEx network,  and the clean speech from target speaker.}
\label{fig:spec}
\end{center}
% \vspace{-25pt}
\end{figure}

As an example, we illustrate the speaker extraction from a female-female mixture speech  by the competitive baseline systems and the proposed SpEx network in Figure \ref{fig:spec}. From the log magnitude spectrum, we observe that the proposed SpEx network outperforms other baseline systems in terms of the recovered signal quality and purification. Some listening examples are available online \footnote{ \url{https://xuchenglin28.github.io/files/taslp2019/index.html}}, of which the first example is the audio illustrated in Figure \ref{fig:spec}.

\subsubsection{Different Gender vs. Same Gender}

Generally speaking, speakers  of the same gender  sound closer than those of different gender. We further report the results of the experiments in Table \ref{tbl:methods_comp} for different and same gender mixture separately. We observe  in   Table \ref{tbl:gender} that the performance of different gender mixture is always better than the same gender. This has been observed in human listening test as reported by Treisman \cite{treisman1964selective} in a behavioural study. It was found that difference in voice (i.e., male versus female) allows more efficient rejection of the irrelevant signal when messages are mixed and played to both ears (i.e., diotic).

From Table \ref{tbl:gender}, we also observe that the proposed SpEx network achieves $34.9\%$ and $40.9\%$ relative SDR improvement, and $15.2\%$ and $15.0\%$ relative PESQ improvement over the best baseline, SBF-MTSAL-Concat, for different and same gender conditions.

\begin{table}[t] 
\centering \caption{SDR (dB) and PESQ in a comparative study of different and same gender mixture under open condition.} 
\centerline{
\small
\begin{tabular}{|r|*{4}{c|}}
\hline
\multirow{2}{*}{Methods} & \multicolumn{2}{c|}{SDR} & \multicolumn{2}{c|}{PESQ} \\ \cline{2-5}
 & Diff. & Same & Diff.  & Same  \\
\hline
\hline
Mixture & 2.5 & 2.7 & 2.29 & 2.34 \\
\hline
SBF-IBM \cite{delcroix2018single} & 7.6 & 5.1 & 2.42 & 2.19 \\ 
SBF-MSAL \cite{delcroix2018single} & 12.0 & 6.9 & 2.82 & 2.43 \\ 
SBF-MTSAL \cite{xu2019optimization} & 12.3 & 7.2 & 2.85 & 2.44 \\ 
SBF-MTSAL-Concat \cite{xu2019optimization} & 12.9 & 8.8 & 2.90 & 2.54 \\ \hline
SpEx & \textbf{17.4} & \textbf{12.4} & \textbf{3.34} & \textbf{2.92} \\
\hline
\end{tabular}}
\label{tbl:gender} 
\end{table}

% \vspace{-2.5pt}
\subsubsection{Mixture with Different SNR}
% \vspace{-2.5pt}

It is of interest to investigate how the proposed SpEx network performs for mixture speech of different SNR, where we consider the target speech as the foreground and the interference as the background noise. We train a SpEx network on the dataset that has the SNR range of [0-5] as described in Section \ref{subsec:data}. The same SpEx network has been reported in Tables \ref{tbl:methods_comp} and \ref{tbl:gender}. 

We divide the test set into $3$ SNR groups, namely [0, 1)dB, [1, 3)dB and [3, 5]dB. The results are summarized in Table \ref{tbl:diff_snr}. Without surprise, test data of higher SNR performs better than that of lower SNR. We also observe that the proposed SpEx network achieves $52.9\%$, $39.6\%$ and $30.4\%$ relative SDR improvement over the best baseline system, SBF-MTSAL-Concat, for [0, 1)dB, [1, 3)dB and [3, 5]dB SNR group respectively. Since the SNR of the simulated database is limited from 0dB to 5dB, in the future work, we will investigate various SNR ranges, i.e., from -10dB to 20dB.

\begin{table}[t] 
\centering \caption{SDR (dB) of extracted speech  when we evaluate the same SpEx system on varying duration of reference speech of target speaker at [0, 1)dB, [1, 3)dB, [3, 5]dB.} 
\centerline{
\small
\begin{tabular}{|r|*{3}{c|}}
\hline
\diagbox{Methods}{SNR(dB)} & [0, 1) & [1, 3) & [3, 5]  \\
\hline
\hline
Mixture & 0.7 & 2.0 & 4.2 \\
\hline
SBF-IBM \cite{delcroix2018single} & 4.0 & 5.8 & 8.4 \\ 
SBF-MSAL \cite{delcroix2018single} & 7.1 & 9.2 & 11.3 \\ 
SBF-MTSAL \cite{xu2019optimization} & 7.5 & 9.5 & 11.5 \\ 
SBF-MTSAL-Concat \cite{xu2019optimization} & 8.7 & 10.6 & 12.5 \\ \hline
SpEx & \textbf{13.3} & \textbf{14.8} & \textbf{16.3} \\ 
\hline
\end{tabular}}
\label{tbl:diff_snr} 
\end{table}

\subsubsection{Subjective Evaluation}
% \vspace{-5pt}

Since the SBF-MTSAL-Concat represented the best baseline performance in the objective evaluation, we only conducted an A/B preference test between the proposed SpEx network and the SBF-MTSAL-Concat baseline to evaluate the signal quality and intelligibility in a listening test. We randomly selected $20$ pairs of listening examples, including the original target speaker's reference and two extracted signals for the target speaker by the proposed SpEx network and the best baseline system. We invited a group of $13$ subjects to select their preference according to the quality and intelligibility. The listeners were asked to pay special attention to the amount of perceived distortion and interference from background. For each test, the subject listened to three audios in a group, the reference speech was firstly played, followed by the extracted speech in random order from the two systems. The subject didn't have the information about which speech stemmed from which system. 

\begin{figure}[!t]
\begin{center}
\includegraphics[width=85mm]{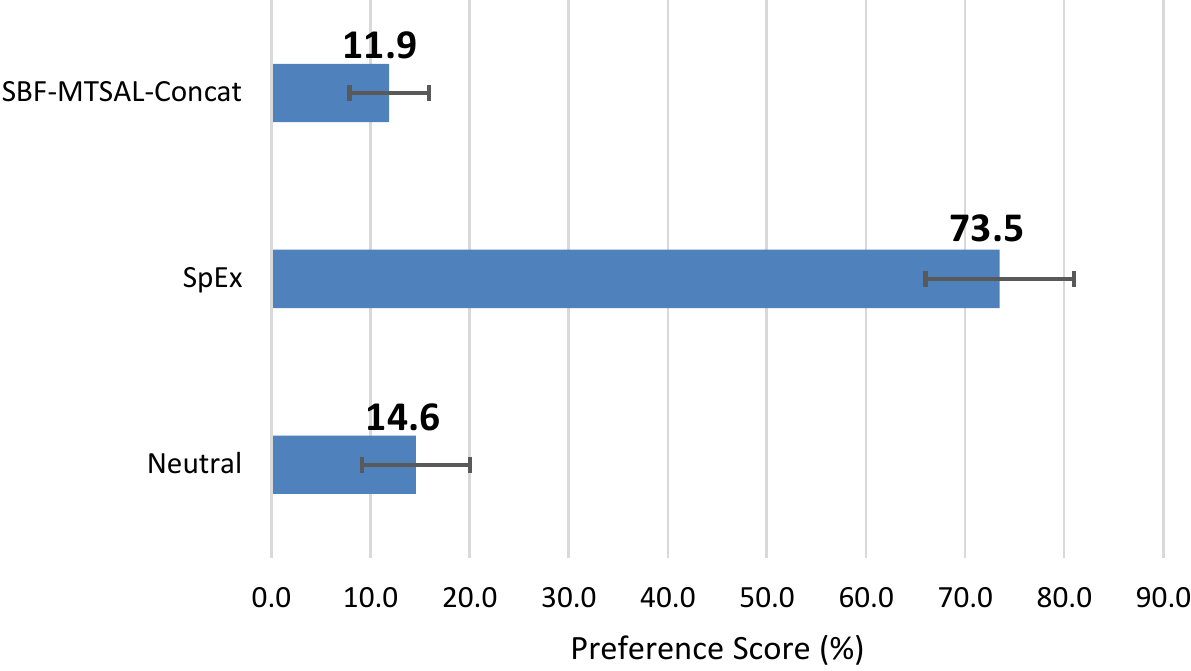}
\caption{The A/B preference test result of the extracted target speaker's voice between the proposed SpEx method and the best SBF-MTSAL-Concat baseline. We conducted t-test using a significance level of $p<0.05$ which is depicted with the error bars.}
\label{fig:ABtest}
\end{center}
% \vspace{-25pt}
\end{figure}

We observe from Figure \ref{fig:ABtest} that the listeners clearly favor the proposed SpEx network with a preference score of $73.5\%$ as opposed to that of $11.9\%$ for the best SBF-MTSAL-Concat system. Most listeners significantly favor the SpEx network with a significance level of $p<0.05$, because of lower distortion and inter-speaker interference than the best baseline.

\subsubsection{Duration of the Reference Speech}
\label{ssec:duration}

As speaker extraction relies on the reference speech of the target speaker to develop the top-down voluntary focus, the duration of the reference speech plays a role in the process. We further look into the impact of the duration on speaker extraction performance. 
In the aforementioned experiments, the duration of the reference speech in training, development and test sets is at ``Random" as described in Section \ref{subsec:data}. Now let's compare the ``Random" setting with different duration groups ($7.5$s, $15$s, $30$s and $60$s) in the test set. The experimental results are summarized in Table \ref{tbl:methods_duration}. 

Since the average duration of the reference speech in the ``Random" condition of the test set is $7.3$s, we firstly evaluate the performance on the test subset with reference speech of a duration $7.5$s. It is noted that the results are similar between ``Random" condition and the $7.5$s subset. When we increase the duration of the reference speech in the test set to $15$s, $30$s and $60$s, we observe that  longer duration leads to  better results in general. When we fix the duration of the reference speech to $15$s for the training and development set, the performance drops slightly when comparing with those under the ``Random" condition. However, we continue to observe that longer test speech duration always helps.

\begin{table}[t]
\centering \caption{ SDR (dB), SI-SDR(dB) and PESQ in a comparative study of different duration of the reference speech. ``Random'' indicates that the duration of the reference speech is random.} 
\centerline{
\footnotesize
\begin{tabular}{|c|*{5}{c|}}
\hline
Training & Test & SDR & SI-SDR  & PESQ  \\
\hline
\hline
\multirow{5}{*}{Random} & Random & 15.1 & 14.6 & 3.14 \\ 
 & 7.5s & 15.0 & 14.6 & 3.14 \\
 & 15s & 15.4 & 15.0 & 3.17 \\ 
 & 30s & 15.5 & 15.2 & 3.19 \\
 & 60s & \textbf{15.6} & \textbf{15.2} & \textbf{3.19} \\
\hline
\multirow{2}{*}{15s} & 15s & 14.9 & 14.5 & 3.13 \\ 
 & 60s & 15.2 & 14.8 & 3.15 \\ 
 \hline
\end{tabular}} %\vspace{-10pt}
\label{tbl:methods_duration}
\end{table}

\subsubsection{Comparisons with Speech Separation on WSJ0-2mix}
\label{sssec:separation}

Most speech separation methods conducted their experiments on the well-known WSJ0-2mix database. To compare with speech separation methods, we trained the proposed SpEx model on WSJ0-2mix database to extract each speaker in the mixture by giving a reference speech of the corresponding speaker. In addition, we re-implemented the Conv-TasNet method \cite{luo2018tasnet_arvix} with the same optimization scheme as our proposed SpEx as described in Section \ref{subsec:data}.

From Table \ref{tbl:comp_separation}, we observe that the proposed SpEx achieves comparable performance as Conv-TasNet \cite{luo2018tasnet_arvix} with the same TCN architecture. While SpEx and Conv-TasNet are comparable in performance, just like other speaker extraction techniques, SpEx offers its unique advantages over other speech separation techniques in real-world applications.

As SpEx relies very much on the quality of the speaker embeddings, we observed that the proposed speaker encoder has outperformed i-vector encoder (refer to Table \ref{tbl:joint_embed}). We will further investigate the performance of SpEx on the speaker database larger than WSJ0-2mix (101 speakers) in the future work.

\begin{table}[t]
\centering \caption{ SDRi (dB), SI-SDR(dB) and PESQ in a comparative study on the WSJ0-2mix dataset under the open condition. ``\#Paras'' refers to the number of parameters of the model. $^\ddagger$ indicates the latest Conv-TasNet with an additional skip-connection in each TCN block. $^*$indicates our re-implementation of the work in \cite{luo2018tasnet_arvix}. For speech separation (SS) task, we report the results evaluated on the oracle-selected streams. For speaker extraction (SE) task, we report the results evaluated on the SpEx-extracted stream.} 
\centerline{
\footnotesize
\begin{tabular}{|c|r|*{5}{c|}}
\hline
Task & Methods & \#Paras & SDRi & SI-SDR  & PESQ  \\
\hline
\hline
\multirow{11}{*}{SS} & DC++ \cite{isik2016single} & 13.6M & - & 10.8 & - \\
 & uPIT-BLSTM-ST \cite{kolbaek2017multitalker} & 92.7M & 10.0 & - & - \\
 & DANet \cite{chen2017deep} & 9.1M & - & 10.5 & - \\
 & cuPIT-Grid-RD \cite{xu2018single} & 53.2M & 10.2 & - & - \\
 & SDC-G-MTL \cite{xu2018shifted} & 53.9M & 10.5 & - & - \\
 & CBLDNN-GAT \cite{li2018cbldnn} & 39.5M & 11.0 & - & - \\
 & Chimera++ \cite{wang2018alternative} & 32.9M & 12.0 & 11.5 & - \\
 & WA-MISI-5 \cite{wang2018end} & 32.9M & 13.1 & 12.6 & - \\ 
\cline{2-6}
 & BLSTM-TasNet \cite{luo2018real} & 23.6M & 13.6 & 13.2 & - \\
 & Conv-TasNet \cite{luo2018tasnet_arvix} & 8.8M & 15.0 & 14.6 & 3.25 \\
 & Conv-TasNet$^\ddagger$ \cite{luo2019conv} & 5.1M & 15.6 & 15.3 & 3.24 \\ 
\hline
SS & Conv-TasNet$^*$ \cite{luo2018tasnet_arvix}  & 8.8M & 14.5 & 14.2 & 3.11 \\
SE & SpEx & 10.8M & 14.6 & 14.2 & 3.14 \\
\hline
\end{tabular}} %\vspace{-10pt}
\label{tbl:comp_separation}
\end{table}

\subsection{Experiments on Three-Speaker Mixture}

The proposed SpEx network has the inherent ability to extract speech from mixture speech of more than two speakers using the same network architecture.  We train the SpEx system under three conditions: only two-speaker mixture data, only three-speaker mixture data, and the combination of two- and three-speaker mixture data. We then evaluate the performance of the trained SpEx systems on two-speaker and three-speaker mixed test data, respectively. From Section \ref{ssec:duration}, we know that the longer duration of the reference speech in the test set achieves better performance. We keep the duration of the reference speech as $15$s and $60$s for a comparison for both two-speaker and three-speaker mixed test data in this experiment.

From Table \ref{tbl:methods_spks}, we observe that the performance of the two-speaker mixture is always better than the three-speaker mixture in the SpEx systems under three conditions with different training data. This is consistent with the findings in a human's performance of a subject evaluation where both listening comprehension and auditory attention decrease significantly as the number of simultaneous audio channels increased \cite{stifelman1994cocktail}. It further confirms that the longer duration of the reference speech achieves better performance. Because the longer duration of the reference speech derives better speaker embedding.

\begin{table}[t]
\centering \caption{ SDR (dB), SI-SDR(dB) and PESQ in a comparative study of different number of speakers in the mixed speech on WSJ0-2mix-extr and WSJ0-3mix-extr datasets. The duration of the reference speech is random during training. ``\#speakers'' indicates the number of speakers in the mixture. ``Dur.'' indicates the duration of the reference speech.} 
\centerline{
\footnotesize
\begin{tabular}{|c|*{6}{c|}}
\hline
\multirow{2}{*}{Training} & \multicolumn{2}{c|}{Test} & \multirow{2}{*}{SDR} & \multirow{2}{*}{SI-SDR} & \multirow{2}{*}{PESQ} \\ \cline{2-3}
& \#speakers & Dur. & & & \\
\hline
\hline
\multirow{2}{*}{2 speakers} & 2 speakers & 15s & 15.4 & 15.0 & 3.17 \\ 
 & 3 speakers & 15s & 5.2 & 5.0 & 2.35 \\ \hline
\multirow{2}{*}{3 speakers} & 2 speakers & 15s & 11.5 & 10.9 & 2.74 \\ 
 & 3 speakers & 15s & 7.9 & 7.3 & 2.40 \\ \hline
\multirow{2}{*}{2 \& 3 speakers} & 2 speakers & 15s & 15.0 & 14.6 & 3.14 \\ 
 & 3 speakers & 15s & 8.9 & 8.4 & 2.54 \\
\hline
\hline
\multirow{2}{*}{2 speakers} & 2 speakers & 60s & 15.6 & 15.2 & 3.19 \\ 
 & 3 speakers & 60s & 5.2 & 5.0 & 2.36 \\ \hline
\multirow{2}{*}{3 speakers} & 2 speakers & 60s & 12.1 & 11.6 & 2.81 \\ 
 & 3 speakers & 60s & 8.3 & 7.8 & 2.44 \\ \hline
\multirow{2}{*}{2 \& 3 speakers} & 2 speakers & 60s & 15.5 & 15.1 & 3.19 \\ 
 & 3 speakers & 60s & 9.1 & 8.7 & 2.57 \\
\hline
\end{tabular}} %\vspace{-10pt}
\label{tbl:methods_spks}
\end{table}

% \vspace{-8pt}
\section{Discussions and Conclusions}
\label{sec:diss}
% \vspace{-5pt}

We propose an end-to-end speaker extraction network (SpEx) that emulates  humans' ability of selective auditory attention. The SpEx network  forms a top-down voluntary focus by using the reference speech of the target speaker. It is particularly useful in cases where speakers are pre-registered to the system, for example, in speaker verification \cite{rao2019target_is} where the target speaker is known to the system through enrollment.

The  SpEx network also overcomes the phase estimation issue in frequency-domain speaker extraction. The improvements are attributed to the dilated convolutional encoder-decoder framework that performs in time-domain, the multi-scale encoding and decoding, and the multi-task learning algorithm. Our experiments show that the SpEx network significantly outperforms the frequency-domain counterparts. 

The ability of human to detect a particular signal from other interference speech or background noise is greatly improved with two ears \cite{arons1992review}. Previous studies \cite{chen2017cracking,wang2018multi} on multi-channel speech separation have shown impressive improvements, particularly in the presence of reverberation and multiple interference speakers. Similarly, we may improve the speaker extraction performance under those adverse conditions by extending the SpEx network for multi-channel inputs, that will be an extension of this work. In addition, SpEx could be extended to enable DPRNN-TasNet \cite{luo2019dual} for speaker extraction by replacing the TCN block with a dual path RNN for improved speech quality.

Humans tend to perceive sounds as coming from locations of visual events \cite{arons1992review}, for example, when we watch television, where an actor’s voice appears to be emanating from his mouth regardless of where the loudspeaker is located. The speaker encoder mechanism in this paper allows for an easy implementation of audio-visual speaker encoder, that will strengthen the top-down voluntary focus in the selective auditory attention.

Brain computer interaction helps connect human brain with assistive devices, i.e., hearing aid device. To assist people with hearing impairment, it would be interesting to study how SpEx can take  non-invasive electro-encephalography (EEG) \cite{o2015attentional} or invasive electro-corticography (ECoG) \cite{han2019speaker} signals, instead of a reference speech, as input to decode the speech from the attended speaker.  

In summary, the proposed SpEx network marks another step towards solving the cocktail party problem. It will potentially improve the performance of many  down-stream speech processing applications, such as speaker verification \cite{rao2019target_is} and speaker diarization.

% or


%\appendices

%\section{}

% % use section* for acknowledgment
% \section*{Acknowledgment}
% To be added.

\ifCLASSOPTIONcaptionsoff
  \newpage
\fi



\bibliographystyle{IEEEtran}
\bibliography{2019taslp}

\end{document}